\documentclass[aps, pre, twocolumn, amsmath, superscriptaddress,showkeys,showpacs]{revtex4-1}
\usepackage[utf8]{inputenc}
\usepackage[title]{appendix}
\usepackage{amsmath}
\usepackage{graphicx}
\usepackage{amssymb,color,tikz,multirow}
\usepackage{mathrsfs}
\usepackage{fontenc}
\usepackage{float}
\usepackage{amsthm}
\usepackage{hyperref}[colorlinks=true,linkcolor=blue,citecolor=blue]
\usepackage{amsfonts}
\usepackage{mathtools}
\begin{document}

\title{How Combined Pairwise and Higher-Order Interactions Shape Transient Dynamics}
\author{Sourin Chatterjee}
\email{sourin.CHATTERJEE@univ-amu.fr}
\affiliation{Department of Mathematics and Statistics, Indian Institute of Science Education and Research, Kolkata, West Bengal 741246, India}
\affiliation{Institut de Neurosciences des Systèmes (INS), UMR1106, Aix-Marseille Université, Marseilles, France}
\author{Sayantan Nag Chowdhury}
\email{jcjeetchowdhury1@gmail.com}
\affiliation{School of Science, Constructor University, 28759 Bremen, Germany}
\affiliation{Physics and Applied Mathematics Unit, Indian Statistical Institute, 203 B. T. Road, Kolkata-700108, India}
\affiliation{Department of Environmental Science and Policy, University of California, Davis, Davis, California 95616, USA}

\begin{abstract}

Understanding how species interactions shape biodiversity is a core challenge in ecology. While much focus has been on long-term stability, there is rising interest in transient dynamics—the short-lived periods when ecosystems respond to disturbances and adjust toward stability. These transitions are crucial for predicting ecosystem reactions and guiding effective conservation. Our study introduces a model that uses convex combinations to blend pairwise and higher-order interactions, offering a more realistic view of natural ecosystems. 
We find pairwise interactions slow the journey to stability, while higher-order interactions speed it up. Employing global stability analysis and numerical simulations, we establish that as the proportion of higher-order interactions (HOIs) increases, mean transient times exhibit a significant reduction, thereby underscoring the essential role of HOIs in enhancing biodiversity stabilization. Our results reveal a robust correlation between the most negative real part of the eigenvalues of the Jacobian matrix associated with the linearized system at the coexistence equilibrium and the mean transient times. This indicates that a more negative leading eigenvalue correlates with accelerated convergence to stable coexistence abundances. This insight is vital for comprehending ecosystem resilience and recovery, emphasizing the key role of HOIs in promoting stabilization. Amid growing interest in transient dynamics and its implications for biodiversity and ecological stability, our study enhances the understanding of how species interactions affect both transient and long-term ecosystem behavior. By addressing a critical gap in ecological theory and offering a practical framework for ecosystem management, our work advances knowledge of transient dynamics, ultimately informing effective conservation strategies.


\end{abstract}

\maketitle

\section{Introduction}

\par Understanding species interactions and their impact on biodiversity remains a pivotal issue in ecological research \cite{lotka1920analytical, volterra1926variations, may1974biological, may1975nonlinear, may2019stability}. Ecology has traditionally focused on studying stable, long-term dynamics in ecosystems, but there is growing recognition of the importance of shorter, temporary periods known as transient dynamics \cite{hastings2001transient,hastings2018transient,lai2011transient}. These are the intervals when an ecosystem is transitioning from its initial state to a potential long-term stable state. These periods are essential for understanding how ecosystems react to changes like the introduction of new species, environmental shifts, or disturbances. Studying transient dynamics helps us predict whether an ecosystem will eventually stabilize to a steady state or continue to fluctuate, providing insights into its resilience and ability to recover \cite{biggs2009turning}. For example, during these transients, we can observe how quickly an invasive species spreads \cite{ludwig1978qualitative} or how an ecosystem responds to a sudden environmental change \cite{hutchinson1961paradox,arnoldi2016resilience,cuddington2011legacy,kuussaari2009extinction}. This knowledge is vital for effective ecosystem management, helping set realistic goals for conservation and restoration, and identifying critical moments for intervention \cite{ray2021mitigating}.

\par However, studying these short-term dynamics presents challenges because current mathematical theories are not yet well-developed, making predictions and management more difficult. Nevertheless, research is increasingly focused on understanding the causes and impacts of transient dynamics and finding ways to manage them. Examples include disease outbreaks \cite{tao2021transient}, shifts in species populations, and significant changes in ecosystems, such as coral reef degradation \cite{norstrom2009alternative}. Early studies have also explored related concepts, such as cyclical plant succession and species' behavior in lakes and savannas under varying conditions \cite{van2003effects,scheffer1993alternative}. Recognizing the importance of transient dynamics is crucial because these temporary periods can lead to unexpected changes in ecosystems that might not be predicted by focusing solely on long-term behavior \cite{hastings2018transient,morozov2020long}. Understanding these dynamics is especially important when managing invasive species and in conservation and restoration efforts \cite{hobbs2001restoration,neubert2000demography,hastings2004transients,shea2002community,hastings2006persistence}, where setting realistic recovery timelines and identifying critical points for intervention are essential. Additionally, transient dynamics reveal complex interactions between species, such as predator-prey relationships and competition, that may be less apparent once the ecosystem reaches a stable state \cite{scheffer2005cascading,murdoch2003consumer,hastings2001transient,schreiber2008crossing,holt2008theoretical}.

\par The importance and complexity of transient dynamics in physical systems have also recently gained attention. There have been reports of transient chaotic behavior in dynamical systems \cite{dudkowski2022extreme}, even in cases where the attractor is hidden within the phase space \cite{nag2020hidden}. In fact, it may be possible that a dynamical system may become temporarily trapped at a local maximum or minimum within a potential landscape for a long time, giving the impression of stability. However, a minor disturbance—such as noise or an external force—can enable the system to overcome this barrier and transition to the lowest energy (or most stable) configuration \cite{koch2024ghost,chowdhury2020effect,bovier2016metastability,kelso2012multistability}. From a different perspective, one recent study \cite{parastesh2022synchronization} explores the role of higher-order interactions in neuronal dynamics by examining a simplicial complex of neurons, revealing that weak second-order interactions can facilitate synchronization at lower first-order coupling strengths, and that three-body interactions reduce overall synchronization costs compared to pairwise interactions. Similarly, Ref.\ \cite{mehrabbeik2023synchronization} investigates higher-order interactions in a memristive Rulkov model network, using master stability functions to analyze synchronization patterns, and demonstrates that incorporating higher-order interactions lowers the required coupling parameters for synchronization while also showing that larger network sizes enhance synchronization dynamics and facilitate cluster synchronization under specific coupling conditions. Many other intriguing studies on higher-order interactions \cite{alvarez2021evolutionary,pal2024global,ma2024social,anwar2024collective,mirzaei2022synchronization,mehrabbeik2023impact,chatterjee2023novel,vasilyeva2021multilayer,alvarez2022collective,majhi2022dynamics,kumar2021evolution} exist; however, most of them primarily emphasis on long-term behaviors.

\par On the other hand, traditionally, models have focused on either pairwise interactions \cite{ray2022extreme,nag2024cluster,ray2020aging,nag2023interlayer,chatterjee2023effective,chowdhury2019convergence} or higher-order interactions (HOI) \cite{battiston2020networks,gibbs2022coexistence,zhang2023higher,battiston2021physics} exclusively. However, real-world ecosystems are likely to exhibit a mix of these interaction types, with species interactions containing varying fractions of higher-order components. Such mixed-interaction frameworks are essential to better reflect the complexities of natural systems and provide more accurate insights into their dynamics. Recent research has focused on identifying HOIs from ecological data and making HOIs more common in ecological systems \cite{terry2020identifying, li2021beyond, raj2024structure, kleinhesselink2022detecting, barbosa2023experimental}. In classical niche theory \cite{chase2009ecological, soberon2009niches}, species coexistence within the same niche is often challenged by competitive exclusion, leading weaker species to be out-competed by stronger ones. Conversely, neutral theory \cite{brokaw2000niche} posits that species are ecologically equivalent, with biodiversity emerging from a stochastic balance between speciation and extinction. While both theories have significantly advanced our understanding, they also have notable limitations. For instance, niche theory struggles with the paradox of the plankton, where the number of coexisting species exceeds the number of available limiting resources \cite{huisman1999biodiversity}. Similarly, neutral theory's ``neutral drift" is incompatible with the observed stability in forest diversity \cite{clark2003stability}.

\par Recent developments in ecological modeling have introduced game theory-based approaches \cite{chowdhury2023eco,park2017emergence,nagatani2019metapopulation,chowdhury2021complex,roy2022eco,roy2023time,nagatani2019infection,ahmadi2023dynamics,chowdhury2021eco, yang2023evolutionary}, such as intransitive cyclic competition models like the rock-paper-scissors (RPS) game, which maintain biodiversity through cyclical dominance. These models demonstrate that even in hierarchical systems, intransitive competition can stabilize species coexistence, a phenomenon known as the ``stabilizing effect of intransitivities." Empirical evidence supports the prevalence of intransitive competition in various ecological communities, such as bacterial strains \cite{kirkup2004antibiotic, laird2006competitive, hibbing2010bacterial, szolnoki2014cyclic, park2017emergence, bhattacharyya2020mortality, islam2022effect, chatterjee2023response} and phytoplankton \cite{soliveres2015intransitive} and parasite-grass-forb \cite{cameron2009parasite}. Although research on systems that consider both pairwise and higher-order interactions is still emerging, it is gaining attention \cite{gibbs2022coexistence,malizia2024reconstructing,kundu2022higher}. One recent study \cite{van2024modification} examined how the speed at which higher-order interactions emerge affects the stability and evolution of ecological networks, offering new insights into this underexplored area. Another study \cite{gibbs2024can} found that while higher-order interactions can lead to equilibrium in species abundance, they do not always ensure stable coexistence. However, when weak or cooperative pairwise interactions align with higher-order interactions, they can foster robust coexistence in diverse ecosystems. Using the generalized Lotka-Volterra model, Ref.\ \cite{singh2021higher} derive a rule showing that while negative higher-order interactions can stabilize species coexistence by strengthening intraspecific competition, positive higher-order interactions can do so across a wider range of conditions by alleviating pairwise competition, and their results extend to multispecies communities, emphasizing the role of negative intraspecific HOIs in maintaining diversity. Our focus differs as we seek to better understand transient dynamics and explore how they respond to the simultaneous presence of both pairwise and higher-order interactions.




\par In this current study, we develop a generalized model having combined interactions from pairwise and higher-order interaction to mimic the real-world scenario in case of species modeling. While this approach does not affect the solution but it affects its stability dynamics and how fast or slow the dynamical system goes to the equilibrium after the transient dynamics.

\begin{figure*}[ht]
    \centering
    \includegraphics[width=\textwidth]{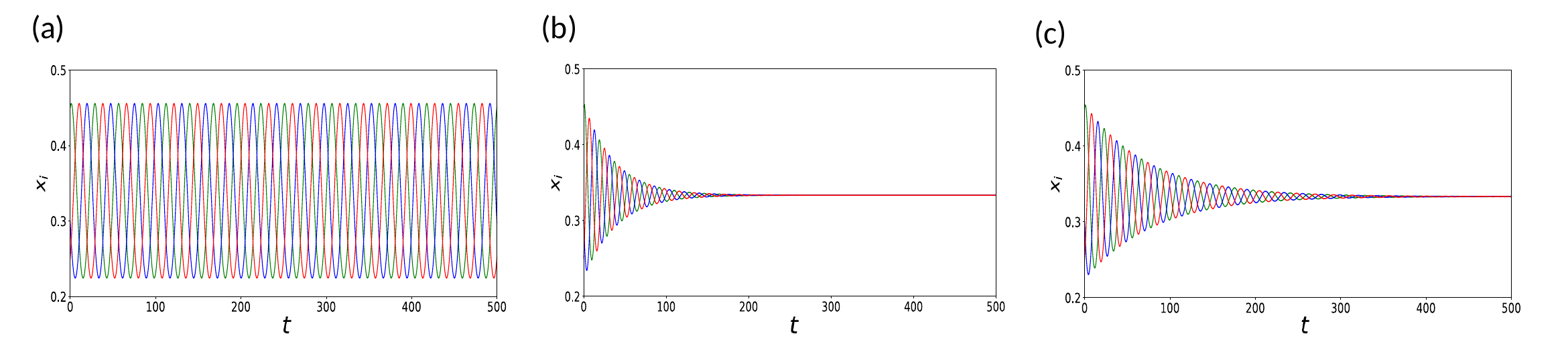}
    \caption{ {\bf The impact of pairwise interactions (\(\gamma_1\)) and higher-order interactions (\(\gamma_2\)) on species coexistence. }Dynamics of model systems with N = 3 for (a) Pairwise $(\gamma_1=1.0)$ (b) Non-pairwise $(\gamma_2=1.0)$ (c) Mixed system $(\gamma_1 \neq 0$, $\gamma_2 \neq 0$, $\gamma_1+\gamma_2=1.0)$. The initial condition
		$(0.45,0.3,0.25)$ is chosen randomly within the $(0,1) \times (0,1) \times (0,1)$ maintaining the constraint $x_1(0)+x_2(0)+x_3(0)=1.0$. As shown in subfigure (a), the pairwise interaction system does not converge to a steady state. However, introducing higher-order interactions significantly alters this behavior. Interestingly, the system with both pairwise and higher-order interactions converges more slowly compared to the system with only higher-order interactions, although both ultimately reach the same equilibrium point $\big(\frac{1}{3}, \frac{1}{3}, \frac{1}{3}\big)$. $\gamma_1 = \gamma_2 = 0.5$ is set for subfigure (c).
} 
    \label{fig:mixing}
\end{figure*}

\section{Model}

\subsection{Description}

\par Reference \cite{grilli2017higher} establishes a model for higher-order interactions and demonstrates how these can be generated using only pairwise interactions. Now, we consider the dynamical model as convex combinations of pairwise and other $(p-1)$ higher-order interactions describing the temporal
evolution of the density $x_i(t)$ of the i-th species as follows:

\begin{equation}\label{eq2}
\begin{split}
{\dot{x}}_{i} &= x_i \bigg[ \gamma_1 \sum_{j=1}^{N} B_{ij}^1 x_{j} + \gamma_2 \sum_{j=1}^{N} \sum_{k=1}^{N} B_{ijk}^2 x_{j} x_{k} \\
& \quad  + \gamma_3 \sum_{j=1}^{N} \sum_{k=1}^{N} \sum_{l=1}^{N} B_{ijkl}^3 x_{j} x_{k} x_{l} + \cdots p\text{-terms}\bigg],
\end{split}
\end{equation}

where $B_{ij}^1 = A_{ij}-A_{ji}$,  $B_{ijk}^2 = 2A_{ij}A_{ik}-A_{ji}A_{jk}-A_{ki}A_{kj}, \text{ and } B_{ijkl}^3 = 3A_{ij}A_{ik}A_{il}-A_{ji}A_{jk}A_{jl}-A_{ki}A_{kj}A_{kl} - A_{li}A_{lj}A_{lk}  \; \forall i,j=1,2,3,\cdots,N$.


\par Here, \(\gamma_i \in [0,1]\) represents the contribution factor, which indicates the percentage of \(B^i\) that contributes to the total interaction.
So,  

\begin{equation*}
    \sum_{i=1}^{p} \gamma_{i}=1.
\end{equation*}

Since $x_i(t)$ represents the density of the $i$-th species, thus the sum of all species density is one, i.e., 

\begin{equation*}
    \sum_{i=1}^{N} x_{i}(t)=1.
\end{equation*}

$A$ matrix encodes information about interactions between two species. As there are $N$ species, $A$ is an $N \times N$ matrix. $A_{jk}$ term account for the winning probability of $j$-th species over the $k$-th species. Hence, 

\begin{equation*}
    A_{jk}+A_{kj}=1 \hspace{0.2cm} \forall j,k.
\end{equation*}

$A_{ij}A_{jk}$ describes the probability of the $j$-th species beating the $k$-th species and, and then, the $i$-th species winning over the $j$-th species. $A_{jj}$ are 0.5 as they are the interactions between the same species.


\par Starting with a generalized equation that includes both pairwise and higher-order interactions with a death rate of one, we use basic algebraic operations to derive the \(B^1\) and \(B^2\) matrices \cite{grilli2017higher,chatterjee2022controlling}. While accounting for pairwise interaction, two species compete against each other. Now, in the first higher-order interaction, we consider the interactions among the possible triplets, where the two species compete with each other, and the winner plays against the third one. Similarly, the compact forms of $B^3, B^4$, etc., can be generated by considering further higher-order terms.

For the sake of simplicity, let us modify the equation as: $\sum_{j=1}^{N} B_{ij}^1 x_{j} \equiv \mathscr{B}^1, \sum_{j=1}^{N} \sum_{k=1}^{N} B_{ijk}^2 x_{j} x_{k} \equiv \mathscr{B}^2,$ and so on. So, Eq.\ \ref{eq2} becomes,

\begin{equation}
    {\dot{x}}_{i} = x_i \bigg[ \gamma_1 \mathscr{B}^1 + \gamma_2 \mathscr{B}^2 + \cdots + \gamma_p \mathscr{B}^p \bigg].
\end{equation}

\subsection{Solution and Global Stability Analysis}

\par The equilibrium point of this system is entirely determined by the matrix $A$. Instead of solving the differential equations directly, the equilibrium point for systems involving pairwise and first higher interactions can be efficiently obtained by calculating the mixed strategy Nash equilibrium of $A$, as demonstrated in Ref.\ \cite{chatterjee2022controlling}. This approach also extends to other higher-order systems, where the equilibrium point can be derived using the same method. Moreover, when creating new systems through linear combinations of existing ones with identical equilibrium points, the equilibrium point remains unchanged, reflecting the consistency of this approach.



\par Replicator dynamics \cite{schuster1983replicator,roca2009evolutionary,borgers1997learning,chowdhury2021extreme}, a key framework in evolutionary game theory, operates as a conservative system, maintaining a constant total population across zero-sum games \cite{hofbauer1998evolutionary}. To demonstrate this for our specific model, we begin with a general interaction matrix $A$. By explicitly calculating the derivatives $\dot{x}_i$ as per Eq.\ \ref{eq2} and performing the necessary algebraic manipulations, we find that the sum of the derivatives  $\sum_i \dot{x}_i$ equals to zero. This result confirms that the system is indeed conservative, ensuring that the total population $\sum_{i=1}^{N} x_i(t) = 1$ remains invariant over time. Hence, the solution of individual density will belong to $[0,1]$. The system is bounded, ensuring that no species' density can fall below zero or exceed one, which is crucial for ecological models where densities represent proportions or fractions. Additionally, the system consistently produces feasible solutions, indicating that the model is well-designed to keep all state variables (such as species populations or densities) within realistic and meaningful limits.

\par Species coexistence is vital in ecology because it underpins biodiversity and ecosystem stability by revealing how species interact and utilize resources. Understanding these dynamics helps in effective conservation, ecosystem management, and predicting the impacts of environmental changes. The coexistence of species in our model is also possible in the presence of intransitive cycles of competitive dominance \cite{grilli2017higher}. 
In this section, to establish the stability of the system in pairwise, non-pairwise, and mixed interaction, we look at a general coexistence equilibrium point  $(x_1^*, x_2^*, ..., x_N^*)$, where $x_i^* \in (0,1)$ for $i=1,2,\cdots,N$.

\par We also study the temporal dynamics of species to validate our findings. For simulations, we use a circulant matrix:

\begin{equation}
A=\rm{Circ.}\big(0.5, \alpha, 1-\alpha, \cdots, \cdots, \cdots),
\end{equation}

where $\alpha \in [0,1] \setminus \{0.5\}$ and $N$ is odd. A circulant interaction matrix simplifies the modeling of species interactions by capturing regular, symmetrical patterns, making it easier to analyze coexistence and stability. It helps in predicting how cyclic or uniform interactions affect species dynamics and system stability. We further assume $\alpha \neq 0.5$ to prevent biologically implausible scenarios \cite{grilli2017higher}. Setting $\alpha = 0.5$ renders all elements of the interaction matrix $A$ equal to $0.5$. This symmetry implies that each species exerts an identical influence on the others, effectively nullifying any cyclic dominance. Consequently, the species are indistinguishable in terms of competitive interaction, resulting in each having an equal probability of winning, with a $50\%$ chance in every interaction.  
For this matrix, the equilibrium point is $\left(\frac{1}{N}, \frac{1}{N}, \frac{1}{N}, \cdots,  \frac{1}{N}\right)$. Since $A$ represents a winning probability matrix, sums like $\sum_{j} B_{ij}^1$ and $\sum_{j,k} B_{ijk}^2$ are zero, leading to $x_i^* = \frac{1}{N}$. By substituting these values into the equation $\dot{x}_i$, we confirm that $\left.\dot{x}_i \right|_{x_i = x_i^*} = 0$. For example, with $N=3$ and $\alpha = 0.7$, the matrix is $A = \text{Circ.}(0.5, 0.7, 0.3)$. We illustrate the dynamics of our model for various values of $\gamma_1$ and $\gamma_2$ in Fig.\ \eqref{fig:mixing}, using initial densities of $(0.45, 0.3, 0.25)$. In the following section, we will analytically demonstrate that these results hold qualitatively consistent for any initial densities within the range $(0,1) \times (0,1) \times (0,1)$. It is important to note that for $N = 3$, selecting $\gamma_3$ or any higher-order $\gamma_i$ with $i = 4, 5, \cdots$ is not possible.


\begin{figure*}[ht]
    \centering
    \includegraphics[width=\textwidth]{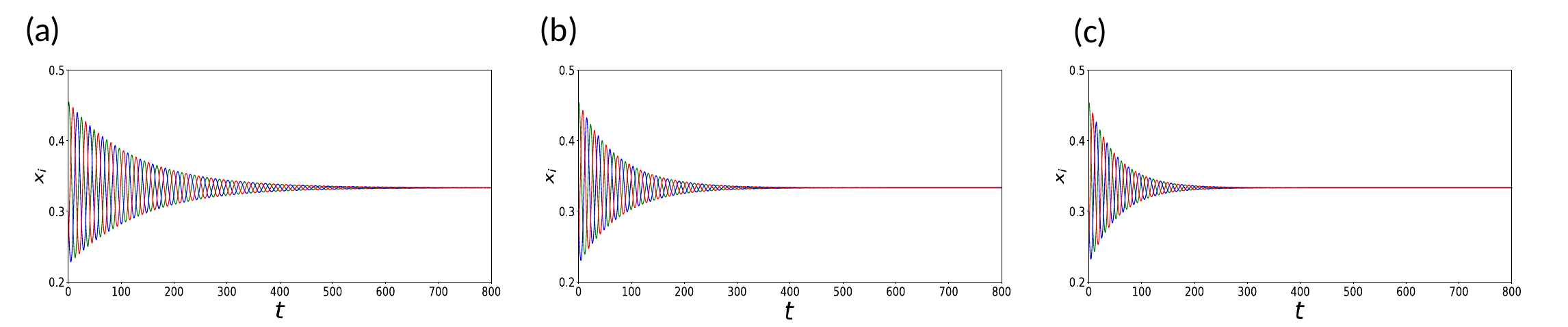}
    \caption{  {\bf The influence of varying proportions of pairwise interactions (\(\gamma_1\)) and higher-order interactions (\(\gamma_2\)) on species dynamics.} Dynamics of mixed systems with $N=3$ for different values of $\gamma_1$ and $\gamma_2$: (a) $\gamma_1 = 0.7, \gamma_2 = 0.3$, (b) $\gamma_1 = 0.5, \gamma_2 = 0.5$, (c) $\gamma_1 = 0.3, \gamma_2 = 0.7$. Starting from the initial condition of $(0.45, 0.3, 0.25)$, as in Fig.\ \ref{fig:mixing}, the system consistently converges to the coexistence equilibrium point, regardless of the relative contributions of pairwise and higher-order interactions. However, as the proportion of pairwise interactions decreases (i.e., as $\gamma_1$ diminishes), the transient time significantly shortens. This demonstrates that higher-order interactions play a crucial role in accelerating convergence towards equilibrium.} 
    \label{fig:mixing2}
\end{figure*}



\subsubsection{Pairwise System}

We consider pairwise interaction model as follows,  

\begin{equation}\label{pair}
{\dot{x}}_{i}=x_i \sum_{j=1}^{N} B_{ij}^1 x_{j}.
\end{equation}

Now, we take the following Lyapunov function \cite{chatterjee2022controlling},
\begin{equation}\label{equ1}
\begin{aligned}
V(x_i)=- \sum_{i=1}^{N} x_{i}^{*}\log\bigg(\frac{x_i}{x_i^*}\bigg),
\end{aligned}
\end{equation}  
where $N \geq 3$ is an odd integer. Now, we use Jensen's inequality \cite{jensen1906fonctions,ruel1999jensen} as $\log$ is a concave function. We find
\begin{equation}\label{equ2}
\begin{aligned}
\sum_{i=1}^{N} x_{i}^{*} \log \bigg(\frac{x_i}{x_{i}^{*}}\bigg) 
\leq \log \bigg( \sum_{i=1}^{N} x_{i}^{*} \bigg(\frac{x_i}{x_{i}^{*}}\bigg) \bigg)\\=\log  \sum_{i=1}^{N} x_i=\log (1)=0.
\end{aligned}
\end{equation}
Thus, clearly $V(x_i) \geq 0$, for $x_i \in (0,1)$ and 
\begin{equation}\label{equ3}
V({x_{i}^{*}})=- \sum_{i=1}^{N} x_{i}^{*}\log\bigg(\frac{x_i^*}{x_i^*}\bigg)=0.
\end{equation}  

Now,

\begin{equation*}
\begin{aligned}
\begin{split}
\frac{dV}{dt} 
= -2 \sum_{i=1}^{N} \sum_{j=1}^{N} x_{i}^{*} A_{ij} x_j + 1\\ 
= - \sum_{j=1}^{N} \big( 2 \sum_{i=1}^{N} x_{i}^{*} A_{ij} \big) x_j + 1 \\
= - \sum_{j=1}^{N} \big( 2 \sum_{i=1}^{N} x_{i}^{*} (1- A_{ji}) \big) x_j + 1\\
= - \sum_{j=1}^{N} \big( 2 \sum_{i=1}^{N} x_{i}^{*} - 2 \sum_{i=1}^{N} x_{i}^{*}A_{ji} \big) x_j + 1\\[\parskip]
\end{split}
\end{aligned}
\end{equation*}

Since $\sum_{j=1}^{N} A_{ij} x_{j}^{*} =\frac{1}{2}$ \cite{grilli2017higher}, we have

\begin{equation*}
\begin{aligned}
\begin{split}  
\frac{dV}{dt}
= - \sum_{j=1}^{N} (2-1) x_j+ 1
= - \sum_{j=1}^{N} x_j+ 1
=0.\\[\parskip]
\end{split}
\end{aligned}
\end{equation*}

\par As a result, the system follows a closed orbit, cycling neutrally around the equilibrium point without ever reaching it unless it begins precisely at that point. As illustrated in Fig.\ \ref{fig:mixing}(a), starting from the initial condition, the species densities oscillate around $\frac{1}{3}$ with a consistent amplitude that remains unchanged over time, indicating neutral cycling around the equilibrium point.


\subsubsection{Non-pairwise System}

We take the simplest possible non-pairwise system as follows,

\begin{equation}\label{nonpair}
{\dot{x}}_{i}=x_i \sum_{j=1}^{N} \sum_{k=1}^{N} B_{ijk}^2 x_{j} x_{k}.
\end{equation}

We choose the same Lyapunov function (Eq.\ \ref{equ1}) for global stability analysis. We take a non-zero perturbation $\eta_j$ such that 

\begin{equation}\label{equ4}
\eta_j=x_j-x_j^*, j=1,2,\cdots,N.
\end{equation} 

Then, $\sum_{j=1}^{N} \eta_j=0$. Now,

\begin{equation*}
\begin{aligned}
\begin{split}
\frac{dV}{dt} 
= -2 \sum_{i=1}^{N} \sum_{j=1}^{N} \sum_{k=1}^{N} x_i^* \big(A_{ij}A_{ik}+A_{ij}A_{jk}\big)x_jx_k + 1\\[\parskip]
=-2 \sum_{j=1}^{N} \sum_{k=1}^{N} \big(\sum_{i=1}^{N} x_i^* A_{ij} \big)A_{jk}x_jx_k \\-2 \sum_{i=1}^{N} x_i^* \big(\sum_{j=1}^{N} A_{ij} x_j\big) \big(\sum_{k=1}^{N} A_{ik} x_k\big) + 1 \\[\parskip]
\end{split}
\end{aligned}
\end{equation*}

Since $\sum_{i=1}^{N} x_i^* A_{ij}= \frac{1}{2}$ and $\sum_{j=1}^{N} \sum_{k=1}^{N} A_{jk}x_jx_k=\frac{1}{2}$, we have

\begin{equation*}
\begin{aligned}
\begin{split}
\frac{dV}{dt}
=-2 \sum_{i=1}^{N} x_i^* \big(\sum_{j=1}^{N} A_{ij} x_j\big)^2+\frac{1}{2}\\
=-2\sum_{i=1}^{N} x_i^* \bigg(\sum_{j=1}^{N} A_{ij} \bigg( x_i^* +\eta_j\bigg)\bigg)^2+\frac{1}{2}\\[\parskip]
=-2 \sum_{i=1}^{N} x_i^* \bigg(\sum_{j=1}^{N} A_{ij} \eta_j\bigg)^2- 2 \sum_{i=1}^{N} x_i^* \sum_{j=1}^{N} A_{ij} \eta_j \\ 
=-2 \sum_{i=1}^{N} x_i^* \bigg(\sum_{j=1}^{N} A_{ij} \eta_j\bigg)^2- \sum_{j=1}^{N} \eta_j \\[\parskip]
=-2 \sum_{i=1}^{N} x_i^* \bigg(\sum_{j=1}^{N} A_{ij} \eta_j\bigg)^2 < 0.
\end{split}
\end{aligned}
\end{equation*}

\par Thus, the system ultimately converges to a globally stable equilibrium point $(x_1^*, x_2^*, ..., x_N^*)$, determined by the interaction matrix $A$. As shown in Fig.\ \ref{fig:mixing}(b), over time, the oscillation amplitude decays, and the species densities gradually stabilize at the equilibrium point $\bigg(\dfrac{1}{3},\dfrac{1}{3},\dfrac{1}{3}\bigg)$, reflecting a balanced coexistence where each species reaches equal abundance.


\subsubsection{Mixed System}


\par Below, we present a biologically motivated model that reflects real-world ecological dynamics by combining two critical components: pairwise interactions, which capture direct species interactions, and higher-order interactions, which account for more complex, multi-species interactions. This hybrid model allows us to explore how such mixed interactions influence ecosystem stability and species coexistence. This model is constructed as a convex combination of these two interaction types, expressed as follows:

\begin{equation}\label{mix}
{\dot{x}}_{i}=x_i \bigg[ \gamma_1 \sum_{j=1}^{N} B_{ij}^1 x_{j} + \gamma_2 \sum_{j=1}^{N} \sum_{k=1}^{N} B_{ijk}^2 x_{j} x_{k} \bigg].
\end{equation}

Now again, we start with the same Lyapunov function (Eq.\ \ref{equ1}), and non-zero perturbation $\eta_j$ (Eq.\ \ref{equ4}) we arrive at,

\begin{equation*}
\begin{aligned}
\frac{dV}{dt}
=  - \gamma_1 \big[2 \sum_{i=1}^{N} x_i^* \sum_{j=1}^{N} A_{ij} x_j + 1\big] \\
- \gamma_2 \bigg[2 \sum_{i=1}^{N} x_i^* \sum_{j=1}^{N} \sum_{k=1}^{N} \big(A_{ij}A_{ik}+A_{ij}A_{jk}\big)x_jx_k + 1\bigg]\\
\end{aligned}
\end{equation*}

Now, using the same simplification as mentioned in the above two analyses, we finally get for $\gamma_2 \neq 0$, 

\begin{equation*}
\begin{aligned}
\frac{dV}{dt}
=  - \gamma_1\big[ 0 \big] - \gamma_2 \bigg[2 \sum_{i=1}^{N} x_i^* \bigg(\sum_{j=1}^{N} A_{ij} \eta_j\bigg)^2 \bigg] < 0. \\
\end{aligned}
\end{equation*}

\par Thus, in this model, the equilibrium point is globally stable, ensuring that regardless of any initial conditions except for a set of measure zero, the system will converge to this point over time. We adapt this definition of global stability from Refs.\ \cite{kassabov2022global,nag2023interlayer}. Biologically, this reflects the stabilization of species densities, where all species reach a balanced state, coexisting at equal proportions. Numerically, as depicted in Figure \ref{fig:mixing}(c), the system stabilizes at the equilibrium point \(\bigg(\dfrac{1}{3},\dfrac{1}{3},\dfrac{1}{3}\bigg)\), confirming the stability of the community dynamics.


\subsection{Transient dynamics}


\par In the pairwise interaction model, the system exhibits infinite transient time, with solutions continuously oscillating around the equilibrium point. However, introducing higher-order interactions significantly reduces this transient period, as observed in purely higher-order interaction models (See Fig.\ \ref{fig:mixing}(b)). Our study will further investigate how incorporating both higher-order and mixed interactions can accelerate the reduction of transient time, leading to faster stabilization.

\begin{figure}[ht]
    \centering
    \includegraphics[scale=0.28]{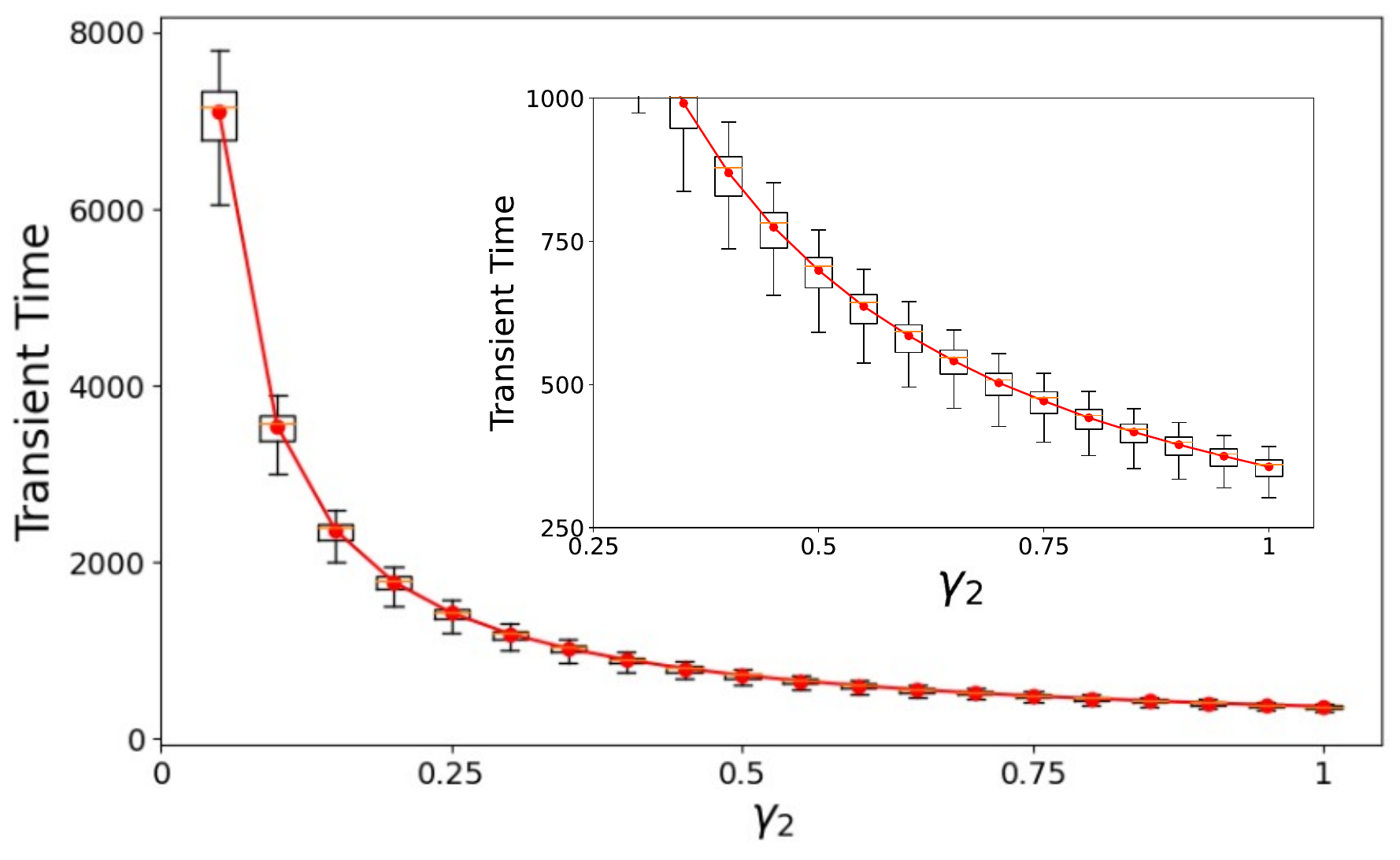}
    \caption{ {\bf Reduction of transient time with increasing higher-order interactions (\(\gamma_2\)).}The boxplot illustrates how transient times vary with different values of $\gamma_2$ across various initial conditions. 
    For the considered interaction matrix $A$, we use ecologically relevant initial conditions and plot the transient times across varying levels of higher-order interactions. For a fixed value of \( \gamma_2 \), we select 20 distinct initial conditions and determine the transient times required to reach the coexistence equilibrium. We then calculate the arithmetic mean of these transient times, represented by the red dots, and the median, indicated by the yellow line. As $\gamma_2$ increases from $0.05$ (indicating a minimal presence of higher-order interactions) to $1.0$ (representing a system solely dominated by higher-order interactions), the transient times steadily decrease. While the higher values of $\gamma_2$ may appear to have minimal variation in transient times due to the large range of the y-axis, the inset provides a clearer view, showing that transient times with different initial conditions are actually well-distributed across this range.}
    \label{fig:tr_ga}
\end{figure}

\begin{figure*}[ht]
    \centering
    \includegraphics[width=\textwidth]{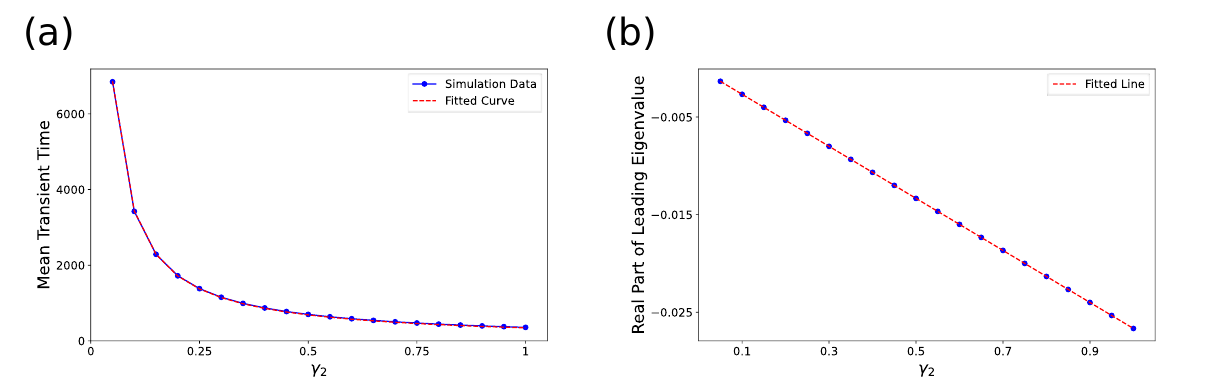}
    \caption{ { \bf Variation of (a) mean transient time and (b) the real part of the leading eigenvalue with respect to \(\gamma_2\). }(a) Mean transient time plotted against the proportion of higher-order interactions $\gamma_2$, with a rectangular hyperbola fit (red dashed line) illustrating how transient time decreases as $\gamma_2$ increases. (b) The most negative real part of the eigenvalues of the Jacobian matrix associated with the linearized system at the coexistence equilibrium plotted against $\gamma_2$ (blue dots), with a linear fit (red dashed line). The negative real part of the leading eigenvalue in (b) provides a direct measure of convergence speed. The correlation between the most negative real part of the leading eigenvalue and the mean transient time illustrates that greater negativity in the leading eigenvalue is associated with faster convergence to the same coexistence equilibrium point. This highlights the role of higher-order interactions in accelerating system stabilization and reducing transient oscillations.}
    \label{fig:tr_ei}
\end{figure*}

\subsubsection{Effects of Shifting Balance from Pairwise to Higher-Order Interactions}

\par 
In Fig.\ \ref{fig:mixing2}, we examine how varying the proportion of pairwise versus higher-order interactions with \(N=3\) affects the system dynamics. Despite different interaction ratios, the system consistently converges to the coexistence equilibrium, aligning with our previous analysis. Notably, as the contribution of higher-order interactions increases and the proportion of pairwise interactions decreases, the system reaches its equilibrium more rapidly. This observation raises a compelling question: {\it why does the system achieve faster convergence with a higher proportion of higher-order interactions (\(\gamma_2 = 1 - \gamma_1\)) and a correspondingly lower proportion of pairwise interactions?} Biologically, this suggests that higher-order interactions may enhance the system’s ability to stabilize by facilitating more complex, cooperative dynamics among species, leading to quicker stabilization of community structure.

\par Before delving deeper into the analytical aspects of our results, we first aim to confirm that incorporating pairwise interactions into our system extends the transient time while increasing higher-order interactions shortens it. Figures \ref{fig:mixing2} and \ref{fig:mixing} illustrate this trend with selected values for parameters $\gamma_1$ and $\gamma_2$. To further validate this observation, we will explore a broader range of data.\\

\subsubsection{Role of Initial Abundances in Systems with a Fixed Interaction Matrix}

\par 
Two approaches can be employed to validate our findings. First, by fixing the interaction matrix \( A \) and varying ecologically meaningful initial conditions, we can analyze how the transient time is influenced by $\gamma_1$ and $\gamma_2$. We perform this analysis with \( A = \text{Circ.}(0.5,0.7,0.3) \) as shown in Figure \ref{fig:tr_ga}. It is well known that when $\gamma_2 = 0$, the system with only pairwise interactions does not converge to the coexistence equilibrium. Thus, we investigate the range $\gamma_2 \in [0.05, 1.00]$ with a small step-length of $\delta\gamma_2=0.05$.

\par In Figure \ref{fig:tr_ga}, transient times are summarized using a boxplot, with the median indicated by the yellow line and the mean by the red dot. Connecting lines between adjacent mean values provide a clearer view of the trend. As $\gamma_2$ increases, indicating a higher proportion of higher-order interactions, mean transient times decrease. This suggests that ecosystems with more complex interactions can stabilize more rapidly, which is crucial for predicting how real-world ecosystems maintain biodiversity. An inset in the figure highlights the reduction in transient time, showing how quickly systems approach stability with higher-order interactions. To create this figure, we selected $20$ different random initial conditions, each constrained by \(x_1(0) + x_2(0) + x_3(0) = 1.0\). We measured the transient time for each condition—defined as the period until density oscillations fall below $10^{-6}$ and then calculated the average of these transient times to obtain the mean value.

\par When $\gamma_2 = 0$, the system's transient time is effectively infinite, leading to a rapid decline near zero and a more gradual decrease as $\gamma_2$ increases. To quantify this relationship, we fitted the mean transient time for $\gamma_2$ values ranging from $0.05$ to $1.0$ with a rectangular hyperbola fit in Fig.\ \ref{fig:tr_ei}(a), illustrating how the product of mean transient time and $\gamma_2$ remains constant, say $c$, as $\gamma_2$ increases. Our statistical analysis suggests that the value of this constant is $c=343.1173$ with a Squared Error Loss of $2705.41$. The Squared Error Loss can be further reduced by increasing the number of initial data points, thereby providing a larger dataset for fitting. This fit provides a statistical measure of how swiftly the system's transient time diminishes as the proportion of higher-order interactions increases. Note that the fitted inverse relationship between \(\gamma_2\) and the mean transient time indicates that as \(\gamma_2\) approaches \(0^+\), the mean transient time tends to infinity, which aligns with our observations. Moreover, the fit demonstrates that the transient duration decreases as \(\gamma_2\) increases, but this reduction occurs in a non-linear fashion. This finding highlights the significant impact of higher-order interactions on accelerating the system's convergence to equilibrium.\\


\subsubsection{Analytical Insights through Linear Theory}

\par 
To further validate our results analytically, we employ linear algebraic theories. We use local stability analysis and examine the leading eigenvalues of the Jacobian matrix  associated with the
linearized system at the coexistence equilibrium. The leading eigenvalue is defined as the eigenvalue with the largest negative real part. The following table compares the leading eigenvalues for the system using the same parameter set as in Fig.\ \ref{fig:mixing2}.


\begin{table}[h!]
\centering
\begin{tabular}{|c|c|}
\hline
\textbf{Parameter} & \textbf{Leading Pair of Eigenvalues} \\
\hline
\(\gamma_1 = 0.7, \gamma_2 = 0.3\) & \(-0.008 \pm 0.2656 i\) \\
\hline
\(\gamma_1 = 0.5, \gamma_2 = 0.5\) & \(-0.0133 \pm 0.2887 i\) \\
\hline
\(\gamma_1 = 0.3, \gamma_2 = 0.7\) &  \(-0.0187 \pm 0.3118 i\) \\
\hline
\end{tabular}
\caption{Comparison of leading pair of eigenvalues}
\label{table:2}
\end{table}

\par The data clearly indicate that as \(\gamma_2\) increases and \(\gamma_1\) decreases, our non-hyperbolic dynamical system exhibits leading eigenvalues with increasingly larger negative real parts. This trend signifies that the system’s rate of convergence to the coexistence equilibrium point accelerates. Larger negative real parts in the leading eigenvalues correspond to a faster decay of perturbations, thereby reducing the time needed for the system to stabilize at the coexistence equilibrium. It's important to note that the equilibrium point remains unchanged regardless of variations in nonzero values of $\gamma_1$ and $\gamma_2$.

\par We have calculated the eigenvalues of the matrix for each value of $\gamma_2$ and plotted the most negative real part in Fig.\ \ref{fig:tr_ei}(b). The results show a clear linear trend with a slope of $-0.0267$ and a zero squared sum error, indicating a precise fit. This linear relationship highlights that as $\gamma_2$ increases, the leading eigenvalues become more negative, which corresponds to faster convergence to the same coexistence equilibrium point. The increased negativity of the leading eigenvalue reflects enhanced stability and quicker decay of transient oscillations, underscoring how higher-order interactions expedite system stabilization.

\par We also compute the Jacobian matrix analytically for a three-species system at the coexistence equilibrium point, parameterized by constants $\alpha$ and $\gamma_2$. Solving for the eigenvalues, we obtain:

\[
\lambda_1 = 0, \quad \lambda_{2,3} = -\gamma_2 \frac{(2\alpha - 1)^2}{6} \pm i \frac{\sqrt{3}}{6} \left(\gamma_2 + 2\right) (2\alpha - 1).
\]

Given that $\gamma_2 \in (0,1]$ and $\alpha \neq 0.5$, the nonzero eigenvalues of the matrix are always a pair of complex conjugates, each possessing both nonzero real and imaginary components. Thus, the real part of the dominant eigenvalue is:

\[
\Re(\lambda_{2,3}) = -\gamma_2 \frac{(2\alpha - 1)^2}{6},
\]

which is always negative irrespective of any $\alpha \in [0,1] \setminus \{0.5\}$ and $\gamma_2 \in (0,1]$. For $\alpha = 0.7$, substituting the values yields \(\Re(\lambda_{2,3}) \approx -0.0267 \gamma_2 \), which is consistent with the negative slope obtained from the linear fit shown in Fig. \ref{fig:tr_ei}.\\


For any linearized dynamical system, the general solution can be expressed as:

$$\sum_{i=1}^{N} C_i \exp(\lambda_i t) v_i,$$

where \(v_i\) are the time-independent eigenvectors and \(\lambda_i\) are the corresponding eigenvalues of the system \cite{murphy2011ordinary}.

Let us define the following:

$$\Re(\lambda_{2,3}) \coloneqq \mathscr{R}, \quad \Im(\lambda_{2}) \coloneqq \mathscr{I},$$

where $\Re(\lambda) \in \mathbb{R}$ and $\Im(\lambda) \in \mathbb{R}$ denote the real and imaginary parts of the eigenvalues, respectively.

In a small $\epsilon$-neighborhood with $\epsilon > 0$ near the equilibrium point, the solution to Eq.\ \ref{eq2}, i.e, the solution to linearised Eq. \ref{eq2} for $N=3$ can be written as:

$$
C_1 v_1 + \exp(\mathscr{R} t)\left[ \Tilde{C_2}\cos(\mathscr{I} t) + i \Tilde{C_3} \sin(\mathscr{I} t) \right],
$$
with $i=\sqrt{-1}$, $\Tilde{C_2} \coloneqq (C_2 v_2 + C_3 v_3)$, $\Tilde{C_3} \coloneqq (C_2 v_2 - C_3 v_3) $ and ${C_i}$'s are time-independent coefficients determined by the initial conditions of the system.

\par Since \(\mathscr{R}\) is negative in our non-hyberbolic system, the term $\exp(\mathscr{R} t)$ represents an exponential decay. 
Although one of the eigenvalues is zero in our case, indicating the corresponding term $C_1 v_1$ neither decays nor grows, which prevents a conclusive determination of linear stability, we establish global stability through the construction of a suitable Lyapunov function. We can assert that the magnitude of \(\mathscr{R}\) governs the rate of convergence towards equilibrium, with the frequency of oscillation determined by the imaginary part $\Im$. However, similar to how linear stability analysis is restricted to the vicinity of the equilibrium, the estimation of the convergence rate based on the real part of the leading eigenvalue \(\mathscr{R}\) is only valid within a local neighborhood of the equilibrium. Nonetheless, our analysis, grounded in the real part of the leading eigenvalues, offers a robust method for estimating the rate at which populations converge to coexistence abundances. Irrespective of the initial abundances, i.e., even if the populations begin at vastly different levels, the system will reliably move toward a balanced state of coexistence, where each species maintains a stable abundance, and our theoretical predictions regarding the rate of convergence towards the coexistence equilibrium align with observed dynamics in the system. This behavior is validated by the results shown in Table \ref{table:2} and Fig.\ \ref{fig:tr_ei}.\\

\begin{figure}[ht]
    \centering
    \includegraphics[scale=0.45]{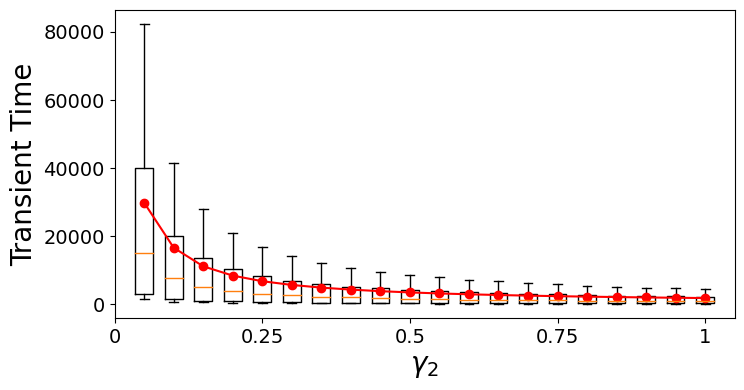}
    \caption{{ \bf The influence of higher-order interactions (\(\gamma_2\)) on transient times across different interaction matrices (\(A\)). }In this analysis, we explore how different interaction structures affect the transient times at varying levels of higher-order interactions. 
 The boxplot shows the variation in transient times for different values of $\gamma_2$ across $20$ randomly generated interaction matrices $A$, with a fixed initial condition $(0.15, 0.35, 0.5)$. The red dots represent mean transient times, while the yellow line marks the median. As $\gamma_2$ increases from $0.05$ (minimal higher-order interactions) to $1.0$ (solely higher-order interactions), transient times decrease, demonstrating the structural robustness of the system in reaching equilibrium. Notably, the coexistence equilibrium remains unchanged across varying $A$, emphasizing that the choice of the particular circulant matrix preserves the coexistence equilibrium and higher-order interactions accelerate convergence to the ecological stability.} 
    \label{fig:tr_A}
\end{figure}

\begin{figure*}[ht]
    \centering
    \includegraphics[width=\textwidth]{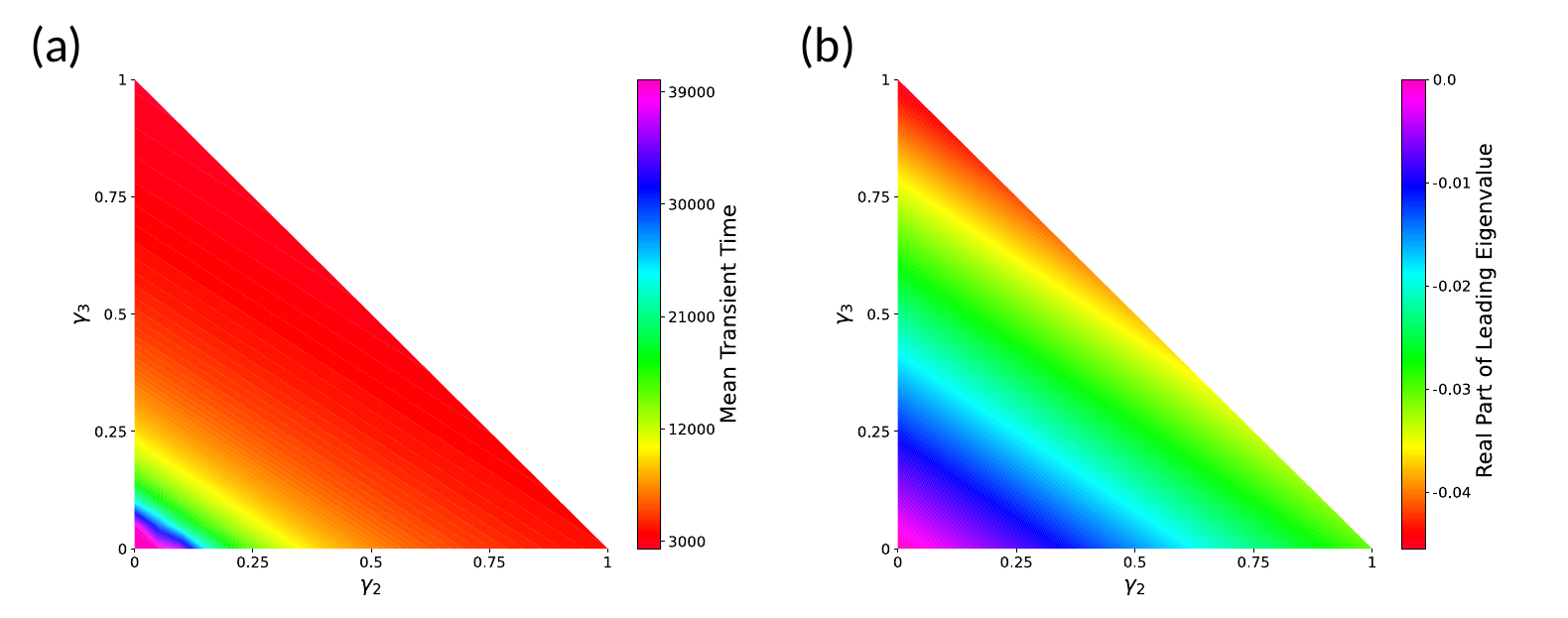}
    \caption{{ \bf The impact of first-order (\(\gamma_2\)) and second-order (\(\gamma_3\)) higher-order interactions (HOIs) on transient times. } (a) The subfigure illustrates how transient time varies with different values of $\gamma_2 \in [0,1]$ and $\gamma_3 \in [0,1]$ with $ \gamma_1 + \gamma_2 + \gamma_3 = 1 $, representing the influence of first and second higher-order interactions (HOIs) on system dynamics. The y-axis shows a more pronounced reduction in transient time along the $\gamma_3$ axis compared to the $\gamma_2$ axis, indicating that second HOI have a stronger impact on accelerating convergence to the equilibrium. This result reflects the ecological insight that more complex interactions (higher-order) promote faster stabilization of ecosystems, enhancing their ability to quickly reach a stable coexistence state. The asymmetry observed between $\gamma_2$ and $\gamma_3$ underscores the differential effects of various HOIs on the system’s transient dynamics.
    (b) The subfigure illustrates how the real part of the leading eigenvalue  of the Jacobian matrix associated with the
linearized system at the coexistence equilibrium shifts with varying values of $\gamma_2 \in [0,1]$ and $\gamma_3 \in [0,1]$, constrained by $\gamma_1 + \gamma_2 + \gamma_3 = 1$, capturing the convergence rate near the system's equilibrium. The $\gamma_3$ axis reveals a more significant negative real part of the leading eigenvalue compared to the $\gamma_2$ axis, indicating that second HOI have a stronger influence on accelerating the system’s return to equilibrium. This asymmetry between $\gamma_2$ and $\gamma_3$ reflects the different biological roles these interactions might play in stabilizing species coexistence; for instance, second HOI could more effectively counterbalance competitive imbalances in ecological communities. This result is consistent with subfigure (a), reinforcing the distinct contributions of first- and second HOI to the system's transient dynamics.} 
    \label{fig:ga_ga}
\end{figure*}

\subsubsection{Effects of Varying Interaction Matrices with Fixed Initial Species Abundances}

\par 
Up to this point, we have explored the influence of higher-order interactions by varying the initial conditions while keeping the interaction matrix $A$ fixed. Alternatively, one can take the opposite approach by keeping the initial condition constant and instead varying the interaction matrix $A$. To explore how the structure of interaction impacts transient dynamics, we fix an initial condition randomly at $(0.15, 0.35, 0.5)$, ensuring $x_1(0) + x_2(0) + x_3(0) = 1.0$, and vary the interaction matrix $A$ by randomly sampling 20 values of $\alpha$ from a uniform distribution over $(0, 1) \setminus {0.5}$. For each $\alpha$, we generate $20$ different circulant $A$ matrices for $N = 3$ and compute the mean transient time, which is plotted in Fig.\ \ref{fig:tr_A}. We observe that increasing $\gamma_2$ consistently reduces transient time, confirming the influence of higher-order interactions on faster convergence. Notably, while varying the initial conditions slightly impacts the transient dynamics, the structure of interactions plays a far more dominant role in determining the transient times, as indicated by the significantly higher variance in transient times compared to different initial conditions. This suggests that the ecological architecture, particularly how species interact through both pairwise and higher-order interactions, exerts a stronger influence on system behavior than initial population distributions. Despite the changes in interaction matrices $A$, the coexistence equilibrium remains constant, underscoring the resilience of the equilibrium point to structural variations in the interaction network. Our circulant matrices ensure that ecosystems with varying circulant interaction patterns will still converge to the same stable coexistence equilibrium. This highlights the role of symmetry and uniformity in maintaining stable ecological balances across diverse interaction networks. However, it should be noted that higher-order interactions accelerate convergence to equilibrium regardless of the interaction matrix structure \cite{grilli2017higher}. 




\subsubsection{Exploring Beyond First Higher Order Interactions}

\par 
In previous analyses, we have examined systems incorporating only pairwise interactions and first higher-order interactions (HOIs). We now extend this investigation to include the effects of additional higher-order interactions. As more HOIs are introduced into the model, we observe that the transient time decreases and the system quickly converges to its equilibrium point. To analyze this phenomenon, we consider a model with pairwise interactions, first HOI, and second HOI, represented by the parameters $\gamma_1$, $\gamma_2$, and $\gamma_3$, respectively. This model adheres to the constraint:

$$ \gamma_1 + \gamma_2 + \gamma_3 = 1. $$

\par By varying $\gamma_2$ and $\gamma_3$, $\gamma_1$ is automatically determined by the constraint, meaning $\gamma_1 = 0$ when $\gamma_2 + \gamma_3 = 1$. When both $\gamma_2$ and $\gamma_3$ are zero, the system only includes pairwise interactions, resulting in an infinite transient time. Consequently, such scenarios are excluded from Fig.\ \ref{fig:ga_ga}. 

\par In the figure, the y-axis shows a more rapid decrease in transient time compared to the x-axis, indicating that second HOI significantly accelerate the system's convergence relative to first HOI. Additionally, the gradient lines in the colormap are not parallel to the line $\gamma_2 + \gamma_3 = 1$, suggesting an asymmetry between the effects of $\gamma_2$ and $\gamma_3$. This asymmetry highlights that second order interactions play a more influential role in reducing transient time than first order interactions, reflecting a more complex and nuanced impact of higher-order interactions on the dynamics of the system.\\

\par Now we analytically compute the Jacobian matrix for $N = 5$, incorporating both first $(\gamma_2)$ and second $(\gamma_3)$ higher-order interactions (HOIs) alongside pairwise interactions $(\gamma_1)$. Although the eigenvalues derived from this analysis are complicated and difficult to express explicitly, we gain better insight into their behavior by plotting them in a two-dimensional triangular space defined by the constraints $\gamma_2 = 0$, $\gamma_3 = 0$, and $\gamma_2 + \gamma_3 = 1$. This analysis aligns with our theoretical expectations; despite the non-hyperbolic nature of the model, the real part of the leading eigenvalue effectively reveals the system's convergence rate toward the coexistence equilibrium. As shown in Fig.\ \ref{fig:ga_ga}(b), there is a clear asymmetry, with $\gamma_3$ exerting a stronger influence on the convergence dynamics than $\gamma_2$. It is important to note that while the transient time decays more rapidly in a non-linear fashion maintaining an inverse relationship with $\gamma_2$, the real part of the leading eigenvalue exhibits a linear trend for a fixed $\gamma_3$. Though both behaviors reflect the rate of convergence toward equilibrium, there is a subtle distinction between the patterns observed in the two subfigures. This behavior is consistent with the earlier observation in Fig.\ \ref{fig:tr_ei}.

\section{Conclusion \& Outlook}

\par While previous studies have typically focused on either pairwise or higher-order interactions in isolation, our approach introduces a model that combines both types of interactions through a convex combination. 
We demonstrate significant differences in behavior between pairwise and higher-order interactions. By incorporating both interaction types, we analytically show that the system converges to a feasible coexistence equilibrium, utilizing appropriate Lyapunov functions for validation. While coexistence theory commonly assumes species interactions are pairwise \cite{levine2017beyond} and much of the literature examines how higher-order interactions (HOIs) affect the stability of species coexistence \cite{letten2019mechanistic, wilson1992complex}, the exploration of transient dynamics has received comparatively less empirical and theoretical attention.
Our research addresses this gap by applying linear theory concepts to assess how HOIs influence transient dynamics in a simple replicator model of a complex multispecies community, considering the presence of pairwise interactions.

\par Our findings reveal that increasing the proportion of higher-order interactions accelerates the convergence to this equilibrium. This is due to higher-order interactions promoting a more efficient stabilization process, whereas pairwise interactions tend to prolong the transient phase, thereby slowing convergence. Although the linearized system with a combination of pairwise and higher-order interactions possesses a zero eigenvalue, the negative real parts of the leading eigenvalues of the Jacobian matrix associated with the linearized system at the coexistence equilibrium indicate that higher-order interactions contribute to faster convergence. This suggests that, biologically, systems with a greater fraction of higher-order interactions are more adept at reaching stable states, whereas pairwise interactions alone may impede this process by increasing transient dynamics.

\par This study employs a simplified replicator model to explore species interactions and their impact on biodiversity, acknowledging that while it does not capture all complexities of real-world systems, it remains valuable for examining various scenarios. Despite extensive research on replicator dynamics, our understanding of transient dynamics is limited due to a lack of established mathematical techniques. We aim to shed light on ecosystem behavior during transient periods, providing insights that may not only enhance our knowledge of ecosystem dynamics but also suggest pathways for more effective management and conservation practices. We focus on a simple replicator model, employing convex combinations to preserve the coexistence equilibrium while comparing transient times. Our model is biologically relevant, as it operates within the interval \([0, 1]\) and reliably converges to the coexistence equilibrium. For simulations, we exclusively use circulant matrices that do not affect the coexistence equilibrium, allowing for meaningful comparisons of transient times. Driven by ecological considerations, we selected this model to explore how different interaction types influence transient dynamics. Our findings highlight the critical role of higher-order interactions in promoting quicker convergence to stable coexistence, which is essential for understanding adaptation and resistance in ecological systems. 

\par Our framework offers fresh insights into natural systems where extended transient phases are prevalent. By integrating both pairwise and higher-order interactions, our model not only deepens the understanding of dynamic stabilization but also suggests new mechanisms for addressing prolonged transient behaviors. These concepts could pave the way for identifying and developing strategies to manage and mitigate such dynamics in ecological and evolutionary systems. We would like to clarify that our findings may not generalize to chaotic or more complex systems, as our analysis is grounded in linear and global stability methods, which are not suited for capturing chaotic dynamics. Likewise, the use of Lyapunov functions is limited to assessing equilibrium stability. We acknowledge that systems exhibiting chaotic behavior require alternative mathematical approaches, which are beyond the scope of this study. Exploring these complexities could offer valuable insights and represents a promising avenue for future research. Another potential avenue for future research is to explore how transient dynamics are affected by time-varying interactions \cite{chowdhury2019synchronization,ansarinasab2024spatial1,chowdhury2019synchronization1,mikaberidze2024consensus,dayani2023optimal,nag2020cooperation,chowdhury2020distance,ansarinasab2024spatial}, a phenomenon commonly observed in natural systems. This remains a central and intriguing direction for future research, with the potential to uncover a broad range of fascinating outcomes.



\section*{acknowledgement}

\par SC gratefully acknowledges Prof.\ Soumitra Banerjee for his deep insights and invaluable feedback during his time at IISER Kolkata, as well as Dr.\ Chittaranjan Hens for valuable discussions during SC’s lab visit to IIIT Hyderabad, funded by IIIT, and for his strong encouragement in pursuing this research. SNC thanks the National Science Foundation (Grant No.\ 1840221) for its support during his postdoctoral tenure at UC Davis, where the problem was first discussed with SC. Further in-depth discussions occurred during the 7th International Conference on Complex Dynamical Systems and Applications 2024 in Digha, India, while SNC was a visiting scientist at the Physics and Applied Mathematics Unit, Indian Statistical Institute, Kolkata. Gratitude is extended to Prof.\ Ulrike Feudel for the opportunity to discuss this work during this period. SNC also thanks Constructor University for its support during his postdoctoral tenure, during which the manuscript's final stages were completed, and Prof. Syamal K.\ Dana for his valuable discussions at Dynamics Days Europe 2024 in Bremen, Germany, and his encouragement. Additionally, SNC expresses deep thanks to Prof.\ Alan Hastings for his insightful discussions at UC Davis and to his wife, Dr.\ Srilena Kundu, for her valuable suggestions, which significantly improved the manuscript.

\bibliographystyle{apsrev4-1}
\bibliography{refs}

\end{document}